\documentclass[prb,amsmath,amssymb]{revtex4}

\usepackage{graphicx}
\usepackage{dcolumn}
\usepackage{bm}

\newcommand\Tm{\langle\mathbf{T}\rangle}

\begin{document}

\title{First-passage times in complex scale invariant media}

\author{S. Condamin}
\affiliation{Laboratoire de Physique Th\'eorique de la Mati\`ere Condens\'ee
(UMR 7600), case courrier 121, Universit\'e Paris 6, 4 Place Jussieu, 75255
Paris Cedex}
\author{O. B\'enichou}
\affiliation{Laboratoire de Physique Th\'eorique de la Mati\`ere Condens\'ee
(UMR 7600), case courrier 121, Universit\'e Paris 6, 4 Place Jussieu, 75255
Paris Cedex}
\author{V.Tejedor}
\affiliation{Laboratoire de Physique Th\'eorique de la Mati\`ere Condens\'ee
(UMR 7600), case courrier 121, Universit\'e Paris 6, 4 Place Jussieu, 75255
Paris Cedex}
\author{R. Voituriez}
\affiliation{Laboratoire de Physique Th\'eorique de la Mati\`ere Condens\'ee
(UMR 7600), case courrier 121, Universit\'e Paris 6, 4 Place Jussieu, 75255
Paris Cedex}
\author{J. Klafter}
\affiliation{School of Chemistry, Tel Aviv University, Tel Aviv 69978, Israel}

\date{\today}

\begin{abstract}

\textbf{ How long does it take  a random walker to reach a given target point?
This quantity, known  as
a first passage time (FPT),  has led to a growing  number of theoretical investigations over the last decade \cite{Redner2001}. The importance of
FPTs originates from the crucial role played by first encounter properties in various real situations, including transport in disordered media \cite{Havlin1987,BenAvraham}, neuron firing dynamics \cite{Tuckwell1988}, spreading of  diseases \cite{Lloyd2001} or target search processes \cite{nousanimaux,nous2D,Shlesinger2006,Eliazar2007}. Most methods to determine the FPT properties in confining domains have been limited to effective 1D geometries, or  for space dimensions larger than one only to homogeneous media \cite{Redner2001}. Here we propose a general theory which allows us to accurately evaluate the mean FPT  (MFPT) in complex media. Remarkably,
this analytical approach provides a universal scaling dependence of the  MFPT on both the volume of the confining domain and the source-target distance. This  analysis is applicable to a broad range of stochastic processes characterized by   length scale invariant properties. Our theoretical predictions are confirmed by numerical simulations for several emblematic models of disordered media  \cite{Bouchaud},  fractals \cite{BenAvraham}, anomalous diffusion \cite{Metzler00} and scale free networks \cite{songpnas}.}
\end{abstract}


\maketitle

Transport properties are often characterized by the  exit time from a sphere $t_{\rm exit}$, which is the first time a random walker reaches
{\it any} point at a distance $r$ from its starting point. This quantity is well known for  Brownian motion in euclidean spaces, and has also been evaluated for finitely ramified deterministic fractals \cite{VandenBroeck, Yuste1995}. In these cases, the length scale invariant properties of the walker's trajectories play a key role and lead to the scaling form $t_{\rm exit}\propto r^{d_w} $, which defines the walk dimension $d_w$. Interestingly, it has been shown very recently that a large class of complex scale free  networks are also invariant under a length scale  renormalization scheme defined in ref.(\cite{songnature}), even if they are of small world type, namely  if their diameter scales like the logarithm of the volume. This remarkable property led in particular the authors of ref.(\cite{songpnas}) to characterize the mean exit time in this class of small world networks by a set of scaling exponents.

However, in many situations, the determining quantity is not $t_{\rm exit}$ but rather the FPT  of a random walk starting from a source point $S$ to a {\it given} target point $T$. Indeed the FPT is a key quantity to quantify the kinetics of transport limited reactions \cite{Rice1985,Yuste1995}, which encompass not only chemical or biochemical reactions\cite{Berg,nousprotein}, but also at larger scales interactions involving more complex organisms, such as a virus infecting a cell\cite{Holcman2007} or  animals searching for food\cite{nousanimaux}. The relevance of the FPT has also been recently  highlighted in ref.(\cite{songpnas}) in the context of scale free networks,  such as social networks \cite{Barabasi1999}, protein interaction networks \cite{Han2004} or metabolic networks   \cite{Almaas2004}.
The FPT and the exit time actually possess very different properties. Indeed, the exit time is not sensitive to the confinment, since only a sphere
of radius $r$ is explored by the random walker. On the contrary,  an estimation of the time needed to go from one point to another, namely the FPT, crucially depends on the confining environment --  the MFPT being actually  infinite in unbounded domains.
Here we propose a general theory which provides explicitly the scaling dependence of the MFPT on both the volume of the confining domain and the source-target distance.

Consider a random walker
moving in a bounded domain of size $N$.
Let $W({\bf r},t|{\bf r}')$ be the propagator, i.e. the  probability density to be
at site ${\bf r}$ at time $t$, starting from the site ${\bf r}'$ at time $0$, and  $P({\bf r},t|{\bf r}')$ the probability density that the first-passage time
to reach  ${\bf r}$, starting from ${\bf r}'$, is $t$. These two probability densities are known to be related through \cite{Hughes}
\begin{equation}
W({\bf r}_T,t|{\bf r}_S) = \int_0^t P({\bf r}_T,t'|{\bf r}_S)
W({\bf r}_T,t-t'|{\bf r}_T) dt'.
\end{equation}
After integration over $t$, this equation gives an exact expression for the MFPT, provided it is
finite:
\begin{equation}\label{Tm}
\Tm = \frac{H({\bf r}_T|{\bf r}_T)-H({\bf r}_T|{\bf r}_S)}{W_{\rm stat}({\bf r}_T)},
\end{equation}
where
\begin{equation}
H({\bf r}|{\bf r}') = \int_{0}^{\infty} (W({\bf r},t|{\bf r}') - W_{\rm stat}({\bf r}))
dt,
\end{equation}
and $W_{\rm stat}$ is the stationary probability distribution (see Supplementary Information (SI) for details). Equation \ref{Tm} is an extension of an analogous form given in ref.(\cite{Noh}), where  no  quantitative determination of the MFPT could be proposed.
The main problem at this stage is to determine the unknown function $H$, which is indeed a complicated task, since it depends both on the walk's characteristics and on the shape of the domain. A crucial step  which allows us to
go further in the general case is that $H$ turns out  to be the pseudo-Green function of the domain \cite{BartonBook}, which in turn is well suited to a quantitative analysis. Indeed, we propose to approximate $H$  by its infinite-space limit, which is precisely the usual Green function $G_{0}$:
\begin{equation}\label{integral}
H({\bf r}|{\bf r}')\approx G_0({\bf r}|{\bf r}') = \int_0^\infty W_0({\bf r},t|{\bf r}') dt,
\end{equation}
where $W_0$ is the infinite space propagator (SI).  Note that a similar  approximation has  proven to be satisfactory in the standard example
of  regular diffusion \cite{Condamin2005a}.
We stress that when inserted in equation \ref{Tm}, this form does not lead to a severe infinite space approximation of the MFPT, since all the dependence  on the domain geometry is
now contained  in the factor $1/W_{\rm stat}$. This approximation is the key step of our derivation and, as  we proceed to show,  captures extremely well the confining effects on MFPTs in complex media.

We first consider the case of a \emph{uniform} stationary distribution $W_{\rm stat}=1/N$, which is realized as soon as the network is undirected and the number of connected neighbors of a node, the degree, is constant. This assumption actually underlies many models of transport in complex media, with the notable
exception of scale free networks, which will be tackled later on.
Following ref.(\cite{BenAvraham}),
we assume for $W_0 $ the standard scaling  :
\begin{equation}
W_0({\bf r},t|{\bf r}') \sim t^{-d_f/d_w} \Pi\left(
\frac{|{\bf r}-{\bf r}'|}{t^{1/d_w}}\right),
\label{scaling}
\end{equation}
 where the fractal dimension $d_f$ characterizes the
number of sites $N_r \sim r^{d_f}$ within a sphere of radius $r$ and $d_w$ has been defined previously. This form ensures the normalization of $W_0$ by integration over the whole fractal set.

 A derivation  given in SI then yields our central result:
\begin{equation}
\Tm \sim \left\{
\begin{array}{ll}
N(A - B r^{d_w-d_f}) & \; {\rm for}\; d_w<d_f\\
N(A + B \ln r) & \; {\rm for}\; d_w=d_f\\
N(A+B r^{d_w-d_f}) & \; {\rm for}\; d_w>d_f
\end{array}
\right.
\label{scalingt}
\end{equation}
for $r=|{\bf r}_T-{\bf r}_S|$ different from $0$.
Strikingly, the constants $A$ and $B$ do not depend on the confining
domain. In addition, while $A$ is related to the small scale properties of the walk, we underline that $B$ can be written   solely in terms of the  infinite space scaling function $\Pi$ (a precise definition of $A$ and $B$ is given in SI). These expressions therefore unveil a universal scaling dependence of the MFPT on the geometrical parameters $N$ and $r$.

Several comments are in order. First, we point out that equation (\ref{scalingt}) gives the large $N$ asymptotics of the MFPT as a function of $N$ and $r$ as \emph{independent} variables. In particular  the volume dependence is linear with $N$ \emph{for $r$ fixed} in any case, which can not be inferred from the standard scaling $\Tm \propto L^{d_w} $, $L$ being the characteristic length of the domain of order $N^{1/d_f}$. However, a global rescaling of the problem $r\to\lambda r,L\to\lambda L$, when applied to equation (\ref{scalingt}),
gives the standard form $\Tm\sim\lambda^{d_w}$ for $d_w>d_f$ and  $\Tm\sim\lambda^{d_f}$ for $d_w<d_f$ in accord
 with ref.(\cite{Montroll1969,Bollt2005}). Second, equation (\ref{scalingt})
  shows two regimes, which  rely on infinite space properties of the walk:
in the case of
compact exploration \cite{BenAvraham} ($d_w\ge d_f$)  where each site is eventually visited, the MFPT behaves like    $\Tm\propto N r^{d_w-d_f}$ ($\Tm\propto N \ln r$ for $d_w = d_f$) at large
 distance, so that
the dependence on the starting point always matters; in the opposite case of non-compact exploration  $\Tm$ tends to a finite value for  large $r$, and the dependence on
the starting point  is lost.

\begin{figure}[ht]
\begin{minipage}{.49\linewidth}
\centering\includegraphics[width = .8\linewidth,clip]{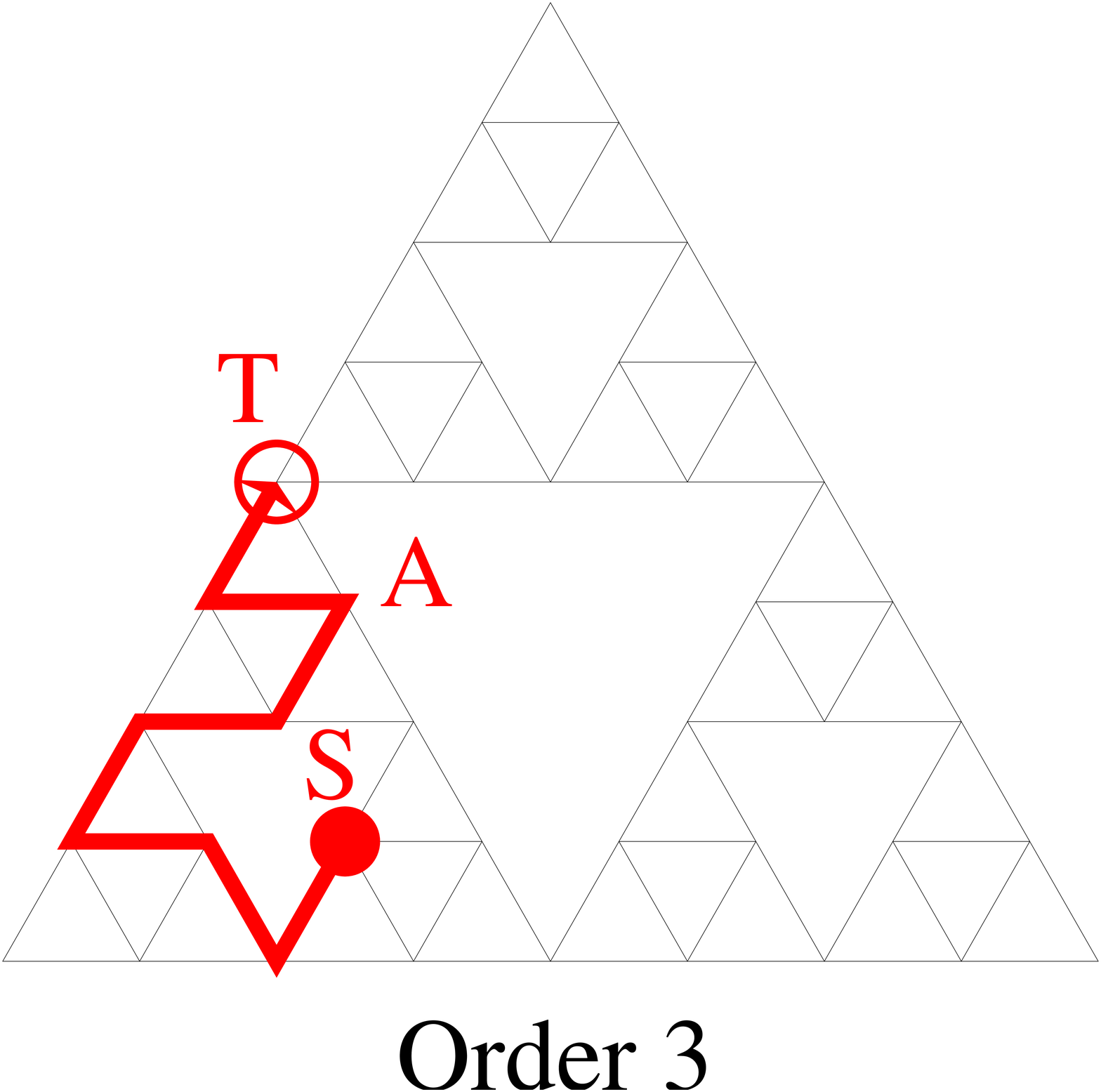}
\end{minipage}
\begin{minipage}{.49\linewidth}
\centering\includegraphics[width = 1.2\linewidth,clip]{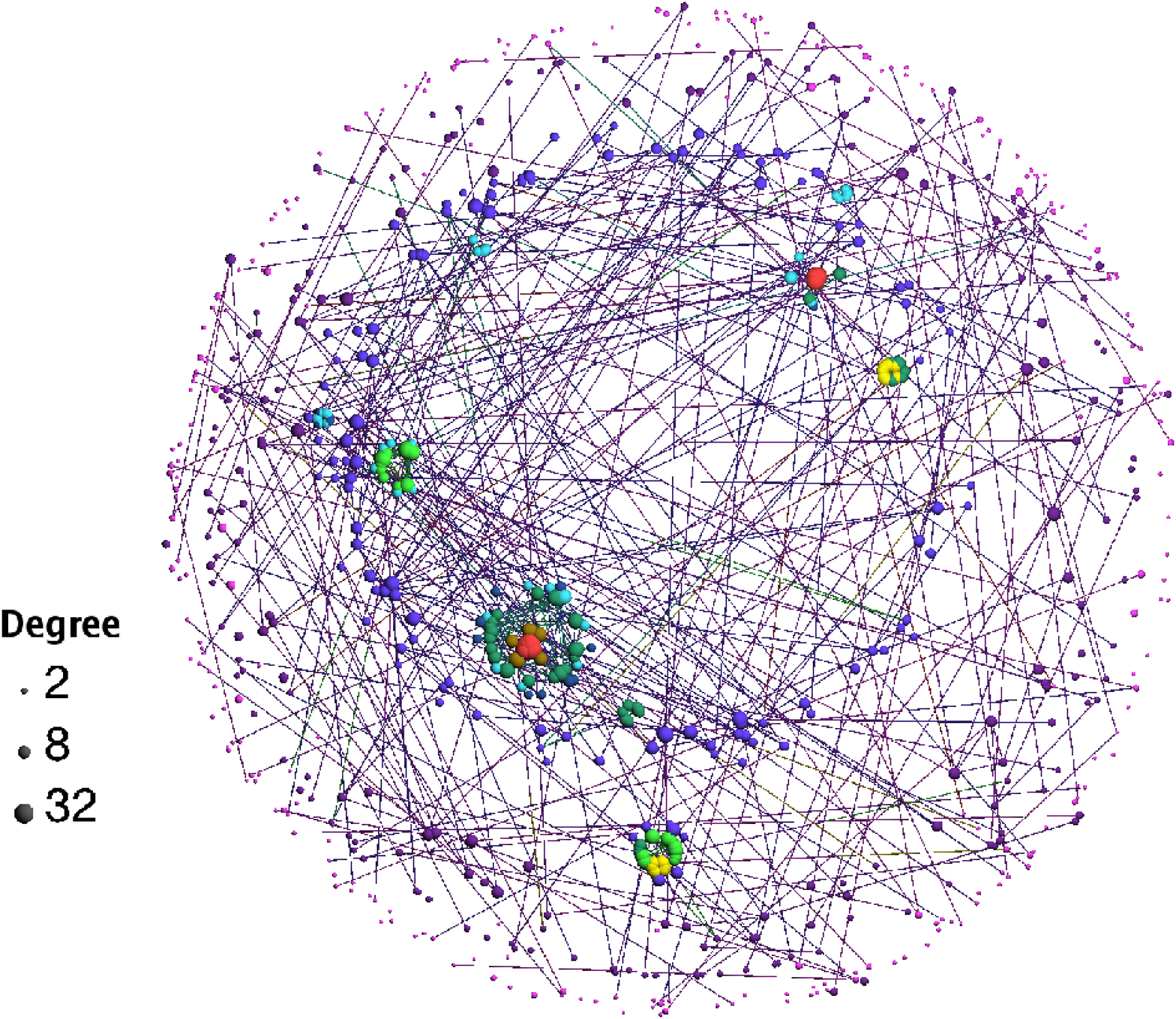}
\end{minipage}
\caption{}
\label{sierpinskipicture}
\end{figure}

We now confirm these analytical results by Monte Carlo simulations and exact enumeration methods applied to various models which exemplify the three previous cases. (i) The random barrier model in 2 dimensions \cite{BenAvraham} is  a widespread model of transport in disordered systems whose MFPT properties remain widely unexplored. It is defined by a lattice random walk with nearest neighbors symmetrical transition rates $\Gamma$  distributed according to some distribution
$\rho(\Gamma)$. Even for a power law distribution $\rho(\Gamma)$ the scaling
function $\Pi(\xi)$ can be shown  to be  Gaussian \cite{Bouchaud} ($d_f=d_w=2$), which allows us to explicitely compute the constant $B$ and obtain $\Tm \sim N\left(A + (1/2\pi D_{\rm eff}) \ln r\right)$. Here   $D_{\rm eff}$ is a diffusion constant depending on $\rho(\Gamma)$ which can be determined by an effective medium approximation \cite{Bouchaud} (SI). (ii)  The Sierpinski
gasket of finite order is a representative example of deterministic fractals, described in Fig. \ref{sierpinskipicture}. In this case\cite{BenAvraham} $d_f = \ln 3/\ln 2 < \ln 5/\ln 2=d_w$, so that our theory predicts the scaling $\Tm \sim N r^{(\ln 5-\ln 3)/ \ln 2}$. (iii) The  L\'evy flight model of anomalous diffusion
\cite{Metzler00,Hughes}  is  based on a
fat-tailed distribution of jump lengths $p(l)\propto l^{-d-\beta}\ (0<\beta\le 2)$. The walk dimension  is now $d_w = \beta$, while the fractal dimension is the dimension of the Euclidian space $d$ . In dimensions $d \geq 2$, or in 1D when $\beta < 1$, one has $d_f>d_w$ and our theory gives $\Tm \sim N \left(A - B r^{\beta-d}\right)$.

Figures 2, 3 and 4 reveal an excellent quantitative agreement between the analytical predictions and the numerical simulations. Both the volume dependence and the source-target distance dependence are unambiguously captured by our theoretical expressions (\ref{scalingt}), as shown by the data collapse of the numerical simulations. We emphasize that the very different nature of these examples demonstrates that the range of applicability of our approach, which mainly relies on the length scale invariant property  of the infinite space propagator (\ref{scaling}),  is wide.

Remarkably, these analytical results can  be extended to scale free networks. The latter  are characterized by a power-law degree distribution. A wide class of scale free networks has been proven recently to be invariant under a length scale renormalization scheme defined in \cite{songnature}: social networks\cite{Barabasi1999}, the world wide web\cite{Albert1999}, metabolic networks\cite{Almaas2004}, and yeast protein interaction networks (PIN) \cite{Han2004}.  Although the standard fractal dimension $d_f$  of these networks is infinite, as their diameter scales as $\ln N$, one can show that they are scale invariant in the following sense : they can  be covered with $N_B$ non overlapping boxes of size $l_B$ with $N_B/N\sim l_B^{-d_B}$. This renormalization property defines an alternative scaling exponent called the box dimension $d_B$, which is actually equal to the fractal dimension defined earlier as long as the networks are not of small-world type. A model of scale free !
 networks possessing such length scale invariant properties has been defined recursively in ref.(\cite{songnaturephysics,songpnas}): the network grows by adding $m$ new offspring nodes to each existing network node, resulting in well defined modules. In addition, modules are connected to each other through $x$ random links (SI). In this case $d_B = \ln(2m+x)/\ln 3$ and $d_w = \ln (6m/x+3)/\ln 3$.

For this class of networks, $W_{\rm stat}({\bf r})$ is not uniform anymore but proportional to the degree $k({\bf r})$ of the node ${\bf r}$. One can use the length scale invariant property to infer the following scaling of the infinite space propagator:
\begin{equation}
\frac{W_0({\bf r},t|{\bf r}')}{k({\bf r})} \sim t^{-d_B/d_w} \Pi\left(
\frac{|{\bf r}-{\bf r}'|}{t^{1/d_B}}\right).
\label{scaling2}
\end{equation}
 This form, compatible with the symmetry relations proposed in ref(\cite{Noh}), allows us to perform a similar derivation which leads for the MFPT to the same result (\ref{scalingt}),  but where $d_f$ is to be replaced by $d_B$. We applied this formula to an example of scale free biological network, the yeast PIN (see figure 1, right), obtained from the filtered yeast interactome developed in  ref.(\cite{Han2004}), and to the model of ref.(\cite{songnaturephysics,songpnas}) of scale free fractal network.  Figure \ref{complex} shows that this analytical result is in good agreement with numerical simulations on the PIN network. The data collapse over various system sizes for the model of scale free fractal networks  provides a further validation of our approach, and indicates that our theory has a wide range of applications.


\newpage

FIG. 1: Length scale invariant networks. {\it Left:} The Sierpinski gasket (here of order 3) is a representative example of deterministic fractal.  A sample random path from $S$ to $T$ is shown. {\it Right:} The yeast PIN, obtained from the filtered yeast interactome developed by  (\cite{Han2004}). Picture generated by the LaNet-vi software  (http://xavier.informatics.indiana.edu/lanet-vi/)

FIG. 2: Random barrier model with a transition rate distribution $\rho(\Gamma)= (\alpha/\Gamma)(\Gamma/\Gamma_0)^\alpha$, with $\Gamma_0 = 1$ and
$\alpha = 0.5$. The confining domain is
a $L \times L$ square with the target point in the middle.
Numerical simulations  of the MFPT rescaled by the volume $N$, averaged over the disorder, for 3 different domain sizes. The theoretical curve  (black line)
is given by  $\Tm/N \sim \left(A + (1/2\pi D_{\rm eff}) \ln r\right)$, where the only fitting parameter is $A$, since $D_{\rm eff}$ is evaluated in SI.

FIG. 3: Numerical simulations of random walks on a Sierpinski gasket (log/log plot) for 3 different system sizes (order 6,7 and 8).
For each set of points, the size of the Sierpinski gasket and the target point
are fixed (the  target point correspond to the point T on the Sierpinski gasket of order 3 in fig. \ref{sierpinskipicture})
, and the starting point takes various positions on the Sierpinski
gasket. The plain line corresponds to the theoretical scaling $r^{d_w-d_f}$.

FIG. 4: Simulations of L\'evy flights on a 2D square lattice  ($\beta=1$).
The confining domains are a $50\times50$ , $100\times100$ and $200\times200$  squares, with the target approximately
in the middle. The MFPT is presented as a function of the
source-target distance for different source points. Simulation points are
fitted with $\Tm \sim N \left(A - B r^{\beta-2}\right)$.

FIG. 5: Simulations of random walks on  fractal complex networks of small world type. The MFPT on the PIN network  (plain blue circles) is fitted by $\Tm/N\sim  \left(A + B r^{d_w-d_B}\right)$, with $d_w\approx 2.86$ and $d_B\approx 2.2$ as found in \cite{songpnas}. We also consider 3 examples of the model of networks defined in \cite{songpnas}:  $(m=3,x=1,d_B-d_w=1,\  {\rm red\ symbols\  and\  fitting\  curve})$,  $(m=3,x=2,d_B-d_w=\ln(3/2)/\ln3,\ {\rm violet\ symbols\  and\  fitting\  curve})$ and  $(m=3,x=3,d_B-d_w=0,\  {\rm green\ symbols\  and\  fitting\  curve})$. For each example, the MFPT (rescaled by the network volume $N=(1+x)^k$, with $k=3,4,5$) averaged over the disorder is presented as a function of the
source-target distance for different source points, and fitted by the theoretical expression $\Tm/N \sim  \left(A + B r^{\ln(3/x)/ \ln 3}\right)$. We find quite surprisingly a scaling independent of $m$.

\newpage

\newpage
\begin{figure}[ht]
\centering\includegraphics[width = .7\linewidth,clip]{barrierbidouille}
\caption{}
\label{random05b}
\end{figure}

\newpage
\begin{figure}[ht]
\centering\includegraphics[width = .7\linewidth,clip]{sierpinskibidouille}
\caption{}
\label{sierpinski}
\end{figure}
\newpage
\begin{figure}[ht]
\centering\includegraphics[width = .7\linewidth,clip]{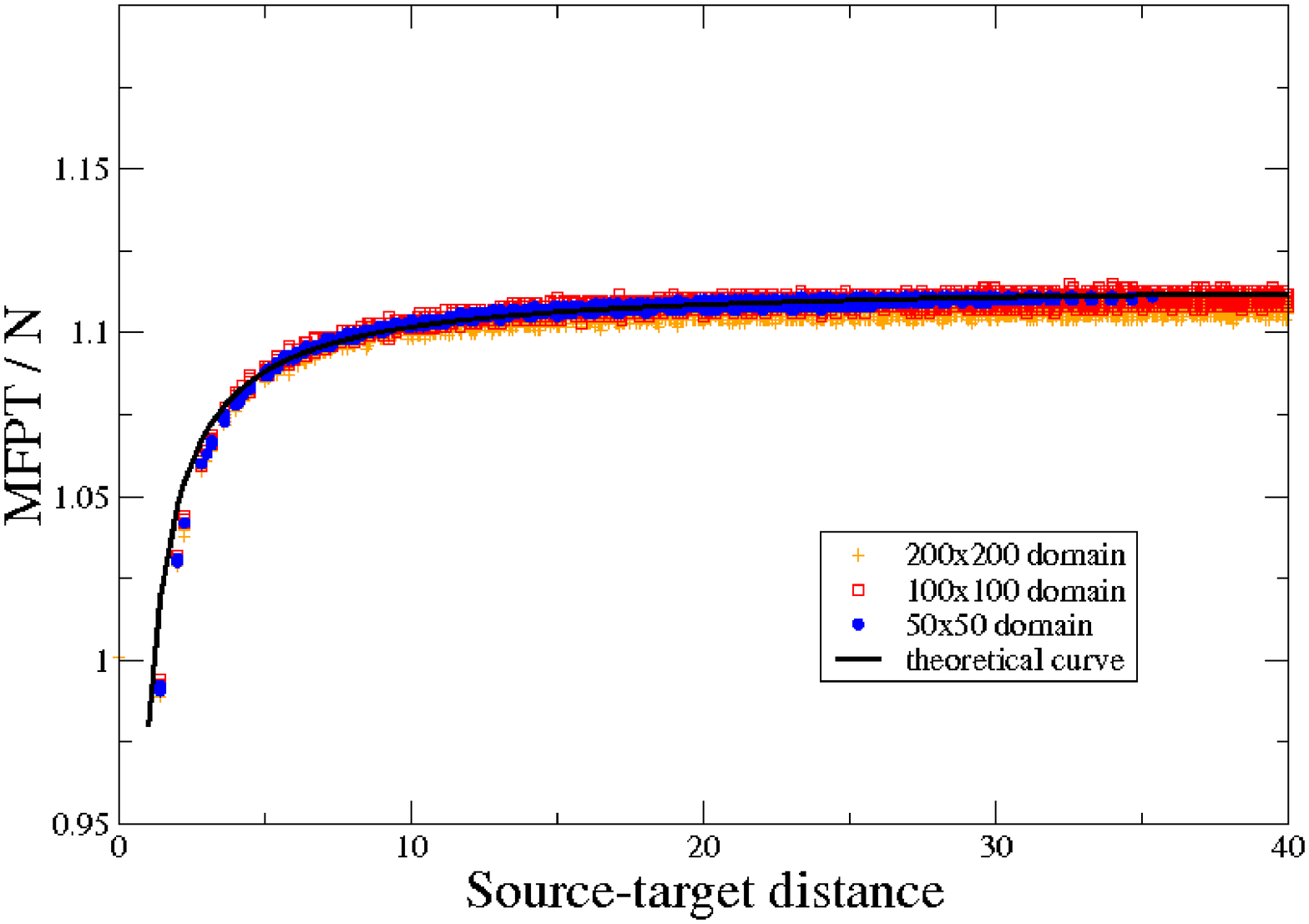}
\caption{}
\label{Levy}
\end{figure}
\newpage
\begin{figure}[ht]
\centering\includegraphics[width = .7\linewidth,clip]{complexbidouille}
\caption{}
\label{complex}
\end{figure}

\end{document}